\documentclass[aps,prb,twocolumn,superscriptaddress, show pacs]{revtex4-2}

\usepackage{graphicx}
\usepackage{dcolumn}
\usepackage{bm}
\usepackage{hyperref}
\usepackage{bm, amsmath}
\usepackage{txfonts}

\begin{document}

\title{Skyrmionium meta-matter: a topologically heterogeneous magnetic crystal with emergent hybrid dynamics.}

\author{Andrey O. Leonov}
\thanks{Corresponding author: leonov@hiroshima-u.ac.jp}
\affiliation{Department of Chemistry, Faculty of Science, Hiroshima University Kagamiyama, Higashi Hiroshima, Hiroshima 739-8526, Japan}
\affiliation{International Institute for Sustainability with Knotted Chiral Meta Matter, Kagamiyama, Higashi Hiroshima, Hiroshima 739-8526, Japan} 

\author{Kaito Nakamura}
\affiliation{Department of Chemistry, Faculty of Science, Hiroshima University Kagamiyama, Higashi Hiroshima, Hiroshima 739-8526, Japan}

\date{\today}

\begin{abstract}
We introduce and systematically investigate a new class of topological magnetic textures—\emph{skyrmionium meta-matter}—composed of skyrmioniums (Skm, $Q\!\approx\!0$) and skyrmions (Sk, $Q\!=\!-1$) arranged in periodic compound lattices that mimic the structural and functional richness of atomic materials. 
While pure skyrmionium lattices (SkmLs) are intrinsically unstable against elongation distortions and relax into the spiral phase, the inclusion of even a small fraction of skyrmions acts as effective topological “pins” that suppress the instability and stabilize a diverse family of mixed Skm–Sk crystals. 
We classify these mixed states by their topological stoichiometry (Skm$_n$Sk$_m$) and demonstrate that each composition can host multiple metastable polymorphs with distinct plane-group symmetries. 
Smooth structural transformations between different polymorphs are achieved by varying the lattice spacing, suggesting possible experimental control via pressure or strain tuning.
The collective spin dynamics of the skyrmionium meta-matter is explored for both in-plane and out-of-plane orientations of the ac magnetic field. 
The resulting absorption spectra exhibit a remarkable richness of resonant modes far beyond the two rotational and one breathing mode characteristic of the conventional skyrmion lattice (SkL). 
We identify hybrid excitations unique to Skm–Sk crystals, including (i) deformation-assisted rotations, in which skyrmions acquire polygonal shapes and rotate around their centers, and (ii) orbital modes, where breathing skyrmioniums induce circular motion of confined skyrmions without altering their size or shape. 
The characteristic frequencies of these collective modes span from sub-GHz to above 10~GHz, consistent with the exchange and DMI energy scales of the system. 
Our results establish skyrmionium-based meta-matter as a versatile platform for realizing tunable, topologically heterogeneous magnetic lattices—an analogue of multicomponent atomic solids— with rich structural and dynamical properties, paving the way toward reconfigurable magnonic and spintronic applications.
\end{abstract}

\maketitle

\section{Introduction}

Topological solitons are a common concept across many areas of physics, appearing naturally in effective field theories that describe systems from nuclear matter to condensed-matter materials~\cite{manton_sutcliffe,shnir,Volovik}. A soliton is, in essence, a self-sustained, particle-like excitation of a continuous field~\cite{solitons}, localized in space and stabilized against dispersion and decay by topological constraints. 
Beyond isolated excitations, solitons can self-organize into extended states, forming periodic arrays or crystal-like structures --- a kind of meta-matter in which they act as quasi-atoms or quasi-molecules.
This self-organization arises from interactions among individual solitons, leading to arrangements that minimize the system's energy. The formation of such extended states is of significant interest as it reveals the interplay between topology, symmetry, and interactions in complex systems and often give rise to emergent phenomena, collective excitations, and novel responses that are absent in single-soliton configurations.

Among many realizations of topological solitons and their meta-matter, two-dimensional (2D) particle-like skyrmions ~\cite{JETP89,Bogdanov94} have become paradigmatic in condensed-matter systems such as chiral magnets (ChM)~\cite{NT}. 

\begin{figure*}
  \centering
  \includegraphics[width=0.99\linewidth]{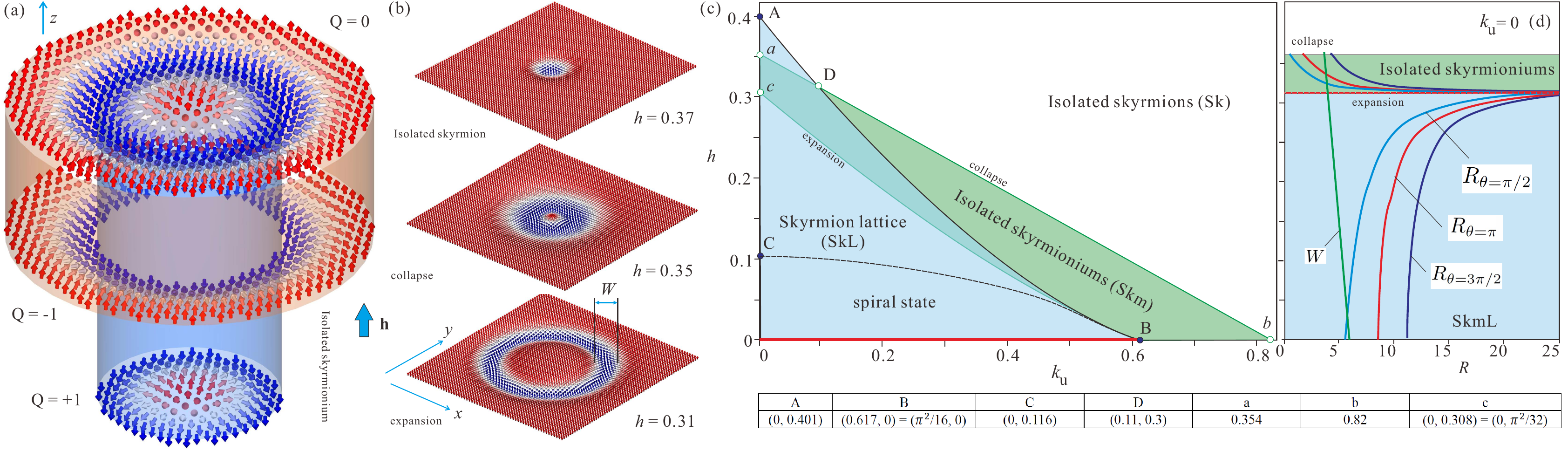}
  \caption{\label{fig01}  
  (a) Schematic of isolated N\'eel skyrmioniums in polar magnets with C$_{nv}$ symmetry or in multilayers with induced DMI. An isolated skyrmionium is a complex object composed of two nested skyrmions with opposite topological charges, $Q=+1$ and $Q=-1$, resulting in a total $Q=0$. 
  (b) Structural evolution of isolated Skms within the field range $c-a$ of the phase diagram shown in (c), exhibiting Skm collapse at point $a$ and expansion at point $c$. 
  (c) Phase diagram of states, highlighting the region of isolated Skms alongside the regions of stable SkL and the spiral phase (see text for details). The red line $0$–$B$ at $h=0$ delineates the region of elliptical instability of the skyrmion lattice. 
 (d) Field-driven evolution of characteristic radii at the levels of the magnetization with $\theta=\pi/2, \pi, 3\pi/2$ for both SkmL and isolated skyrmioniums ($k_u=0$). In general, all radii diverge at the critical line $c - B$. The width of the circular domain wall $W$ [bottom spin configuration in (b)], however, passes without any interruption  over the critical field value. The table summarizes the coordinates of various critical points introduced in the phase diagram and discussed in the text. 
  }
\end{figure*}

Skyrmions are classified by the second homotopy group~\cite{Faddeev,Bott}, $\pi_2(S^2) = \mathbb{Z}$, and carry an integer-valued topological charge, $Q=(1/4\pi)\int \mathbf{m} \cdot \left( \frac{\partial \mathbf{m}}{\partial x} \times \frac{\partial \mathbf{m}}{\partial y} \right) \, dx\, dy$, 
where $\mathbf{m}$ denotes a continuous unit vector field, representing the magnetization in ChM. 
The magnetization at the center of a skyrmion points opposite to the surrounding homogeneous state and rotates smoothly, aligning with it at the skyrmion boundary [Fig. \ref{fig01}]. Thus, the topological charge $Q$ can be interpreted as the degree of wrapping of the unit sphere $S^2$ by the continuous field $\mathbf{m}(\mathbf{x})$. Each value of $\mathbf{m}$ corresponds to a point on $S^2$, and $Q=1$ indicates that the magnetization covers the sphere exactly once as the spatial coordinates $(x,y)$ sweep the plane~\cite{NT,Kovalev2018}.

In ChMs, the characteristic size of a skyrmion~\cite{Bogdanov94,Bogdanov99} is governed by the competition between exchange and Dzyaloshinskii--Moriya interactions (DMI)~\cite{Dz58,Moriya}, and can vary from a few atomic spacings to several micrometers~\cite{Wiesendanger2016}. 

Chiral skyrmions were first observed in bulk cubic helimagnets with intrinsic Dzyaloshinskii–Moriya interaction (DMI), such as MnSi~\cite{Muehlbauer09} and FeGe~\cite{FeGe}. 
In these materials, skyrmions form three-dimensional (3D) filamentary textures aligned along the applied magnetic field~\cite{Damien,Birch:2020} and exhibit the Bloch-type rotational sense.
 Subsequently, versatile thin-film multilayers provided a two-dimensional platform that supports \textit{N\'eel}-type skyrmions.
The breaking of inversion symmetry at the interfaces between a heavy-metal layer and a magnetic layer gives rise to interfacial DMI, as exemplified by PdFe/Ir(111) bilayers~\cite{Romming2013}. These systems offer a wide range of possibilities for selecting magnetic, nonmagnetic, and capping layers, as well as for stacking multiple units to engineer tailored magnetic interactions and skyrmionic properties.

In recent years, magnetic skyrmions have attracted tremendous interest owing to their potential applications in information storage and processing technologies~\cite{Sampaio13,Tomasello14,Shigenaga}. Their unique combination of properties—including topological stability~\cite{Cortes-Ortuno}, nanometer-scale dimensions~\cite{Wiesendanger2016}, and efficient manipulation by low-density electric currents~\cite{Schulz,Jonientz}—makes them highly promising for spintronic device architectures. In particular, the skyrmion racetrack concept~\cite{Wang16,Fert2013} envisions information encoded in individual skyrmions driven along narrow magnetic nanostrips. 

Skyrmionic quasiparticles can also self-organize into extended meta-matter known as skyrmion lattice or skyrmion crystal (SkX). When the eigen-energy of an isolated skyrmion becomes lower than that of the surrounding uniform phase, a spontaneous condensation of skyrmions occurs, giving rise to a periodic array with an equilibrium inter-skyrmion spacing \cite{Bogdanov94}. This mechanism of SkL formation—through the nucleation and subsequent condensation of individual skyrmions—follows the framework introduced by de~Gennes for continuous transitions into incommensurate modulated phases \cite{deGennes}.
Such skyrmion meta-matter provides an attractive platform for magnonics: its collective spin dynamics feature two rotational spin-wave modes (clockwise and counterclockwise) driven by in-plane ac magnetic fields, along with a breathing mode excited by an out-of-plane ac field~\cite{Mochizuki,Desplat}.

Beyond conventional skyrmions, chiral magnetic systems can host more intricate topological solitons in which multiple skyrmionic configurations are concentrically nested within a single structure—so-called $k\pi$ \cite{Bogdanov99} or target skyrmions \cite{Leonov14}. In these composite textures, the magnetization undergoes successive radial rotations, giving rise to alternating skyrmionic and antiskyrmionic rings sharing a common core. The total topological charge varies between $Q=1$ and $Q=0$ depending on whether $k$ is odd or even.

The isolated skyrmionium represents the simplest trivial member of the $k\pi$-skyrmion family ($2\pi$ skyrmion)~\cite{Komineas} and exemplifies the concept of interacting or ``communicating'' skyrmions [Fig.~\ref{fig01}(a)]~\cite{Nakamura}. In this composite structure, the radii of the central skyrmion with positive polarity and topological charge $Q=+1$ [bottom spin configuration in Fig.~\ref{fig01}(a)] and the surrounding ring-shaped skyrmion with negative polarity and $Q=-1$ [middle spin configuration in Fig.~\ref{fig01}(a)] are mutually dependent. For large central skyrmions, the skyrmionium approaches a narrow circular domain wall, whereas for small central skyrmions, the outer ring expands to accommodate the topology [Fig.~\ref{fig01}(b)].

The idea of “a skyrmion within a skyrmion” is particularly attractive
as a way to overcome certain limitations associated with conventional skyrmions. For example, unlike ordinary skyrmions, skyrmioniums possess a vanishing net topological charge ($Q \!\approx\! 0$), as their magnetization field first wraps and then unwraps the unit sphere when the spatial coordinates $(x,y)$ sweep the plane, resulting in a complete cancellation of the topological winding. This property is known to suppress the transverse deflection of the structure, enabling more controlled, rectilinear motion in racetracks. In addition to this topological advantage, skyrmioniums are shown to exhibit enhanced mobility~\cite{Kolesnikov18,Wang20}.

There exists a growing body of experimental work on $k\pi$ skyrmions. 
High-symmetry nanostructures, such as magnetic nanowires~\cite{Higgins}, nanodisks~\cite{Butenko}, and nanorings~\cite{Ponsudana21}, are commonly employed to host them. These geometries stabilize the spin textures through surface effects and provide additional negative energy contributions from edge states, which can favor target skyrmions over other solitonic configurations~\cite{Leonov14}. 
Recently, target skyrmions were realized by weakly coupling 30-nm-thick Permalloy (Ni$_{80}$Fe$_{20}$) disks with 1-$\mu$m diameters to asymmetric (Ir 1 nm/Co 1.5 nm/Pt 1 nm)$\times$7 multilayers exhibiting Dzyaloshinskii--Moriya interaction~\cite{Kent}. Off-axis electron holography further enabled imaging of target skyrmions in 160-nm-diameter nanodisks of the chiral magnet FeGe~\cite{Zheng}.
Skyrmioniums have also been experimentally observed in thin-film geometries without stabilization from lateral boundaries. Notable realizations include thin ferromagnetic films coupled to a magnetic topological insulator~\cite{Zhang}, frustrated Kagome magnets Fe$_3$Sn$_2$~\cite{Yang23}, and flakes of the van der Waals magnet Fe$_{3-x}$GeTe$_2$~\cite{Pwoalla23}.
The stability of Skm has been explored through calculations of the energy barrier separating skyrmionium and skyrmion states using the geodesic nudged elastic band method in Ref. ~\cite{Hagemeister}. More recently, thermal decay and topological transformations of skyrmioniums have been investigated analytically and numerically within micromagnetic frameworks~\cite{Jiang24}, revealing complex pathways of annihilation and conversion into conventional skyrmions.

At the same time, although individual skyrmioniums have been increasingly studied, the physics of skyrmionium meta-matter remains largely unexplored. An open question is whether skyrmionium lattices can form at all, and if so, whether the condensation mechanism of isolated skyrmioniums resembles that of isolated skyrmions. Thus, it is still uncertain which benefits of single skyrmioniums would carry over or be amplified in a SkmL, with potential relevance for magnonics and spintronic devices.

In the present paper, we investigate the stability and spin dynamics of skyrmionium-based meta-matter under the combined influence of an external magnetic field and easy-axis anisotropy.  
Our analysis reveals that a pure skyrmionium lattice is intrinsically unstable, regardless of its symmetry—whether square or hexagonal—and inevitably relaxes into a modulated spiral phase, which represents the true global energy minimum within the same parameter range.  
An indirect signature of the instability boundary is found in the inflation behavior of isolated skyrmioniums: as the system parameters approach this boundary, the eigen-energy of an isolated skyrmionium  tends toward zero but never becomes negative, thus preventing the condensation of skyrmioniums into a stable SkmL.  
A similar instability manifests in the conventional hexagonal skyrmion lattice, which loses its stability at zero magnetic field and continuously elongates into a spiral configuration. This behavior is mirrored in the evolution of isolated skyrmions: as the uniaxial anisotropy decreases, their size diverges  and the eigen-energy also never becomes negative.

Despite this inherent instability of the pure SkmL, the skyrmionium-based meta-matter can attain metastability and form local energy minima when skyrmioniums and skyrmions coexist and intermix to create composite lattices denoted as Skm$_n$Sk$_m$.  
A representative example is the square staggered lattice SkmSk, characterized by the plane group $P4gm$, in which alternating skyrmions and skyrmioniums occupy distinct sublattices.  
Further exploration of other mixed configurations reveals a diversity of realizations with different topological stoichiometries, giving rise to distinct physical properties.  
Such mixed skyrmionium–skyrmion crystals are particularly intriguing from the viewpoint of magnonics, as they host a remarkably rich spectrum of collective spin-wave excitations, including hybrid breathing–rotational modes, deformation-assisted oscillations, and composition-tunable dynamical responses.  
These findings establish skyrmionium meta-matter as a versatile platform for studying the interplay between topological solitons and for engineering dynamical functionalities in chiral magnetic systems.

\begin{figure*}
  \centering
  \includegraphics[width=0.99\linewidth]{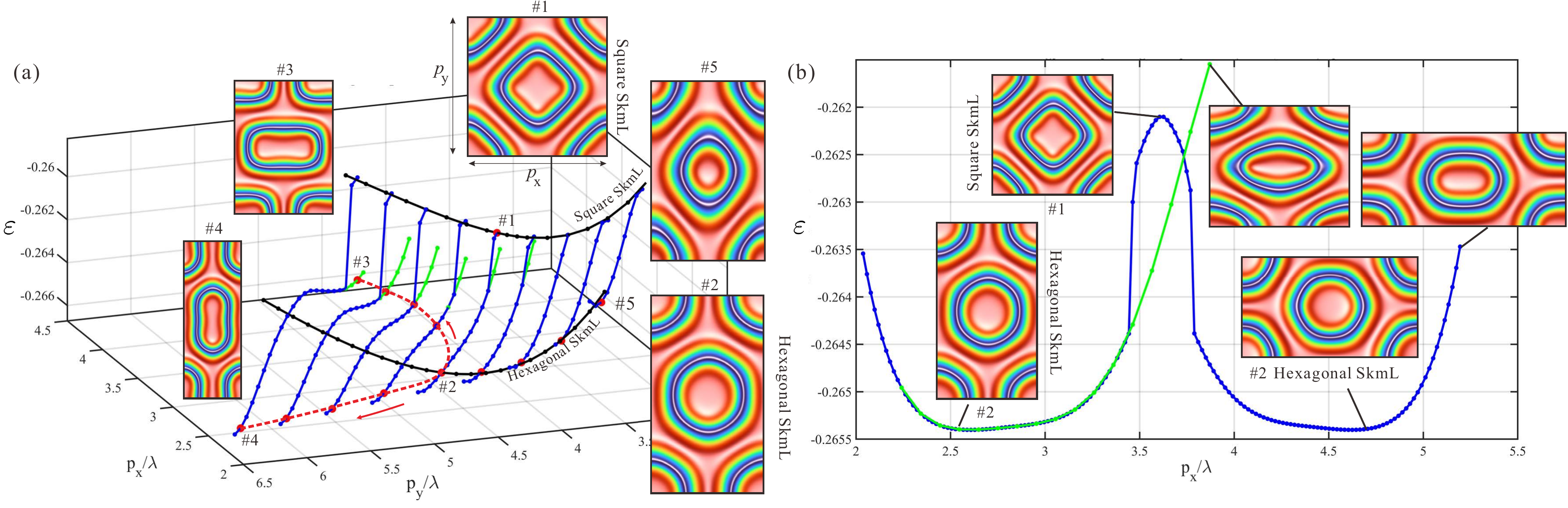}
  \caption{\label{fig02}  
  Instability of the pure skyrmionium lattice. $h=0.2$, $k_u=0$. (a) Energy densities of the square and hexagonal SkmLs, shown as black curves with red dots \#1 and \#2 marking the energy minima. Snapshots \#1 and \#2 illustrate the corresponding magnetization distributions at the energy minima for fixed unit-cell geometries. Blue curves show the energy variation as a function of the unit-cell size, revealing the instability of the hexagonal SkmL under lattice deformation: the SkmL elongates toward configuration \#4 and eventually transforms into a spiral state. The elongation path toward configuration \#3 is separated by a small energy barrier. (b) Evolution of the magnetization patterns along the path connecting the two energy minima corresponding to the hexagonal and square SkmLs (see text for details).
  }
\end{figure*}

\section{Phenomenological theory of skyrmioniums in two-dimensional helimagnets \label{section:phenomenology}}

\subsection{Micromagnetic energy functional\label{sect:model}}
The magnetic energy density of a chiral magnet with C$_{nv}$ symmetry (or with the induced DMI) can be expressed as the sum of exchange, DMI, Zeeman, and anisotropy energy densities:
\begin{equation}
w(\mathbf{m}) = \sum_{i,j} (\partial_i m_j)^2 + w_{DMI} - \mathbf{m}\cdot\mathbf{h} - k_u m_z^2.
\label{functional}
\end{equation}
Spatial coordinates $\mathbf{x}$ are measured in units of the characteristic length of modulated states, $L_D = A/D$. Here, $A > 0$ is the exchange stiffness, $D$ is the Dzyaloshinskii constant, and $k_u = K_u M^2 A/D^2$ is the non-dimensional anisotropy constant. In this work, we restrict ourselves to the easy-axis case, i.e., $k_u > 0$.

For applications to specific material systems, these non-dimensional units can be readily converted to physical units to obtain corresponding skyrmionium solutions. As an instructive example, we consider Co/Pt films, with DMI $D = 3$~mJ/m$^2$, saturation magnetization $M_s = 580$~kA/m, exchange stiffness $A = 15\times10^{-12}$~J/m, 
so that the unit length in the forthcoming figures corresponds to $L_D = 5$~nm.

Alternatively, the length scale can be measured in units of $\lambda$:
\begin{equation}
\lambda=4\pi L_D,
\label{lambda}
\end{equation}
which is the period of the spiral state for zero magnetic field (e.g., 18 nm for the bulk MnSi or 60 nm for Cu$_2$OSeO$_3$ \cite{Seki2012,Crisanti}).

We consider a thin ferromagnetic film in the $xy$ plane with periodic boundary conditions (pbc). The normalized magnetic field is $\mathbf{h} = \mathbf{H}/H_0$, applied along the $z$ axis, where $H_0 = D^2 / (A|\mathbf{M}|)$. The magnetization vector $\mathbf{m}(x,y) = \mathbf{M}/|\mathbf{M}|$ has fixed magnitude normalized to unity.

For magnets with C$_{nv}$ symmetry, the DMI energy density takes the form
\begin{equation}
w_{DMI} = m_x\partial_x m_z - m_z\partial_x m_x + m_y\partial_y m_z - m_z\partial_y m_y,
\label{DMI}
\end{equation}
where $\partial_x = \partial/\partial x$ and $\partial_y = \partial/\partial y$.

As the primary numerical tool to minimize the functional~(\ref{functional}), we employ the \textsc{mumax3} software package (version 3.10), which simulates magnetization dynamics by solving the Landau--Lifshitz equation using a finite-difference discretization technique~\cite{mumax3}.  
To validate the obtained solutions, we cross-checked the results using our own numerical routines based on simulated annealing and single-step Monte Carlo dynamics with the Metropolis algorithm, which are described in detail elsewhere~\cite{metamorphoses} and will not be repeated here.

\subsection{The phase diagram of states}

The phase diagram of equilibrium states for model~(\ref{functional}) in the plane of control parameters $h$ and $k_u$ was previously reported, e.g.,  in Ref.~\cite{Mukai}, and is reproduced here in Fig.~\ref{fig01}(c).

One-dimensional (1D) spiral states occupy the curvilinear region $c$–$B$–$0$. Along the line $c$–$B$, the spiral period diverges, releasing a set of repulsive isolated kinks with positive eigen-energy relative to the homogeneous state. Conversely, for magnetic fields below the line $c$–$B$, isolated 1D kinks acquire negative energy and tend to fill the plane with an equilibrium inter-kink spacing.

2D skyrmion lattices exist in the region $A$–$B$–$0$ and exhibit analogous behavior. Along the line $A$–$B$, the SkL expands, while isolated skyrmions can condense into a lattice at the same line, where their eigen-energy becomes negative relative to the surrounding homogeneous state.

Throughout the entire parameter range, the region of SkL stability lies above that of the spiral state: a first-order transition between these two modulated phases occurs along the line $C$–$B$. 
Notably, the tricritical point reported in earlier studies \cite{Bogdanov94,Bogdanov99,Butenko2} is absent here. In Ref.~\cite{nanomaterials2024}, it was shown that, slightly below the line $A-B$, the hexagonal SkL continuously transforms into a square SkL, thereby extending the overall stability region of the SkL and eliminating the previously assumed critical point.

As shown in Ref.~\cite{Nakamura}, isolated skyrmioniums exist as a distinct branch of solutions within the narrow region $a$–$b$–$B$–$c$ of the phase diagram. Along the line $a$–$b$, a Skm collapses into an ordinary skyrmion, whereas along $c$–$B$, it transforms into a circular domain wall [Fig. \ref{fig01} (b)] and expands up to infinity. In Fig. \ref{fig01} (d), we plot the characteristic sizes of isolated Skms above the field $h(c)$ at some particular levels of the magnetization: $\theta=3\pi/2$ (dark-blue curve, $m_z=0$), $\theta=\pi$ (green curve, $m_z=-1$), and $\theta=\pi/2$ (light-blue curve, $m_z=0$). 

As will be shown later, this behavior implies the inherent instability of skyrmionium lattices in the considered system. In particular, the energy of an expanding isolated Skm never becomes negative with respect to the homogeneous state, preventing its condensation into an extended modulated phase. This contrasts sharply with spiral and skyrmion states, where the eigen-energy of isolated solitons eventually crosses into negative values—signaling the onset of lattice formation while maintaining a finite soliton size. 
Therefore, isolated skyrmioniums cannot act as stable building blocks for a periodic SkmL within this geometry. Consequently, any realization of a (meta)stable SkmL must necessarily be tested within, or in close proximity to, the stability region of the spiral phase.

In practice, when the magnetic field is reduced below the metastability window, isolated skyrmioniums become elliptically unstable and expand to fill the available space, reconstructing the corresponding modulated spiral phase with its equilibrium period. Indeed, below the critical field $h(c\text{–}B)$, the rotational DMI energy within the narrow ring-like domain wall of a skyrmionium dominates, leading to an overall negative eigen-energy. Consequently, the domain wall—whether flat or circular—tends to expand and fill the entire space in the same manner as a simple spiral.

\begin{figure}
  \centering
  \includegraphics[width=0.99\linewidth]{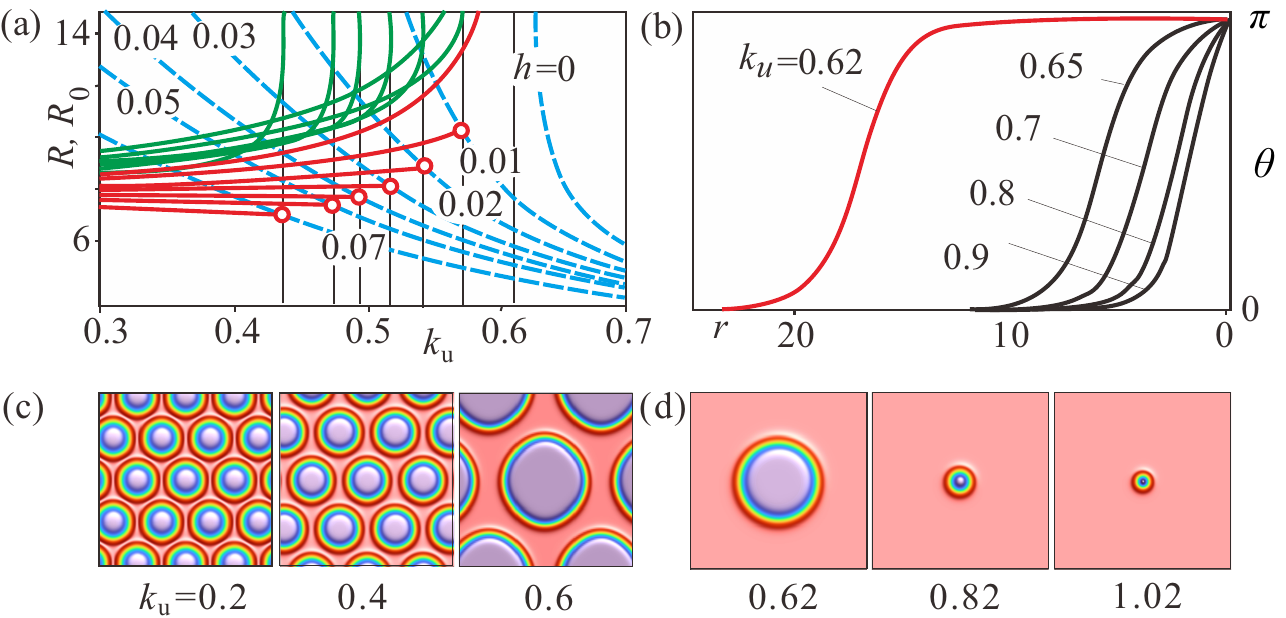}
  \caption{\label{fig03A}  
(a) Expansion of isolated skyrmions and skyrmion lattices as the system approaches the critical anisotropy value $k_u(B)$ from the right and left, respectively. The blue dashed curves show the core size of isolated skyrmions determined according to Lilley’s definition, the red solid lines represent the skyrmion core size within a skyrmion lattice, and the green solid lines indicate the lattice period. For a fixed magnetic field, increasing $k_u$ leads to a critical point where the lattice releases isolated skyrmions. This transition corresponds to the intersection of the red and blue curves, where the lattice period diverges and the skyrmion core size exhibits a discontinuous jump. For $h=0$, the skyrmion lattice and isolated skyrmions remain disconnected.  
(b) Angular profiles $\theta(r)$ of isolated skyrmions in zero magnetic field for different values of the easy-axis uniaxial anisotropy $k_u$, illustrating the progressive expansion of the skyrmion cores as $k_u$ approaches the critical value $k_u(B)=0.61685$ from the right.  
(c), (d) Snapshots showing the expansion of the SkL and isolated skyrmions for different values of the uniaxial anisotropy at $h=0$. The depicted area is the same for all structures and measures $30 \times 30$.  
  }
\end{figure}

\section{Unstable skyrmionium and skyrmion  lattices}

\subsection{Unstable SkmL}

To explore different skyrmionium arrangements in \textsc{mumax3}-simulations, we employ a centered rectangular unit cell containing two skyrmioniums [see snapshots in Fig. \ref{fig02}]. The unit cell dimensions are denoted by $(p_x, p_y)$, where $p_x = N_x \Delta_x$ and $p_y = N_y \Delta_y$. All simulated structures share the same grid resolution of $N_x = N_y = 256$, while the grid spacings $\Delta_x$ and $\Delta_y$ (i.e., the cell sizes in \textsc{mumax3}) are systematically varied to tune the lattice periodicity. In this way, varying the lattice spacings leads to the rearrangement of the constituent isolated skyrmioniums and gives rise to different lattice orders. We remark that the temperature-annealing method used in Ref.~\cite{metamorphoses} also adjusts the lattice spacings $\Delta_x$ and $\Delta_y$, thereby reaching the same energy minima as obtained with \textsc{mumax3}.

First, as an illustrative example, we consider the square lattice [spin configuration \#1 in Fig.~\ref{fig02} (a)], in which the cell sizes are equal, $\Delta_x = \Delta_y$, with control parameters $h=0.2$ and $k_u=0$.
The black line shows the energy density
\begin{equation}
\varepsilon = \frac{1}{V} \int w(x,y) \, dV, \quad  \nonumber
\end{equation}
where $V = N_x N_y N_z \Delta_x \Delta_y \Delta_z$ is the volume of the computational unit cell. Minimization with respect to the period of the square SkmL yields a well-defined energy minimum (marked by the red circle). 

The hexagonal SkmL has a fixed ratio of the cell sizes, $\Delta_x = \sqrt{3}\,\Delta_y$, as shown by the other black line in Fig.~\ref{fig02} (a). It also exhibits a clear energy minimum, marked by the red circle, which corresponds to the equilibrium lattice parameters, spin configuration \#2. In Fig.~\ref{fig01}(d), we quantified the radii corresponding to this hexagonal Skm order, demonstrating its gradual expansion with increasing magnetic field.

The results of energy minimization for the square and hexagonal Skm orders with constrained lattice parameters do not, however, imply that such SkmLs are stable. As the next step, we vary the cell dimensions to connect the energy minima corresponding to the hexagonal and square Skm orders [blue curve in Fig. \ref{fig02} (b)]. Moving along the curve, we employ perfectly circular skyrmioniums as the initial configuration to prevent any artificial ellipticity in the system. 
To achieve this, the relaxation procedure is performed in two stages: first, the magnetic field is set to $h = 0.35$, which promotes circularization of the skyrmioniums within the unit cell without causing their collapse; subsequently, the field is reduced to $h = 0.2$ to obtain the equilibrium state for further analysis.

The resulting energy curve exhibits two energy minima corresponding to rotated hexagonal lattices, while the square Skm order becomes a pronounced energy maximum. Thus, at this stage, we can exclude the square SkmL from consideration, as it would inevitably "roll down" into one of the energy minima with the hexagonal SkmL.

In a similar manner, we plot a set of parallel energy curves connecting the square and hexagonal Skm arrangements [Fig. \ref{fig02}(a)]. It is clearly seen that the energy minimum corresponding to the hexagonal SkmL shifts as the lattice becomes more elongated, leading to the spin configuration~\#4 in Fig.~\ref{fig02}(a). 
This behavior indicates a gradual transformation towards the spiral state. Notably, there is no energy barrier that could prevent this transformation. Interestingly, the hexagonal SkmL can also follow a route toward the elongated spin configuration \#3 (but with small energy barrier). To obtain this trajectory, 
we use the preceeding spin configuration as an initial state along the path [green curves in Fig.~\ref{fig02}(a),(b)].

Thus, our results clearly demonstrate that the SkmL is inherently unstable and tends to elongate into the energetically favorable spiral state. This instability, however, could potentially be suppressed in confined nanostructures that restrict the elongation of the SkmL and help maintain its integrity. 
This behavior underscores a fundamental distinction between skyrmion-based and skyrmionium-based meta-matter: the $Q=0$ topology of skyrmioniums inhibits the formation of stable, periodic lattices.

\begin{figure}
  \centering
  \includegraphics[width=0.99\linewidth]{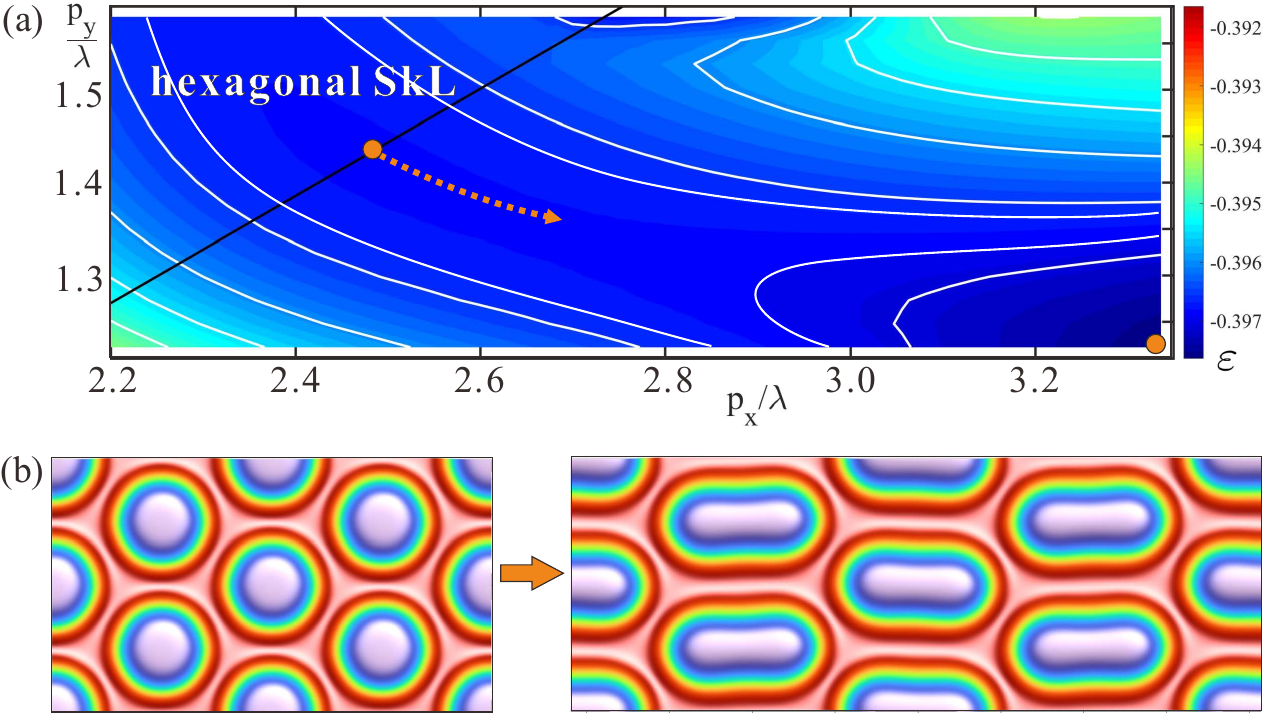}
  \caption{\label{fig03B}  
(a) Energy-density “fingerprint” calculated as a function of the unit-cell dimensions. The black line traces the energy density of the hexagonal SkL for a fixed ratio of lattice parameters. 
The absence of a distinct local energy minimum corresponding to the hexagonal SkL—which becomes discernible only on the two-dimensional energy surface—indicates its instability and tendency to elongate into a spiral state (highlighted by the dotted orange arrow).  
(b) Snapshots illustrating the transition from a perfectly hexagonal skyrmion lattice to the elliptically unstable configuration.
  }
\end{figure}

\begin{figure*}
  \centering
  \includegraphics[width=0.8\linewidth]{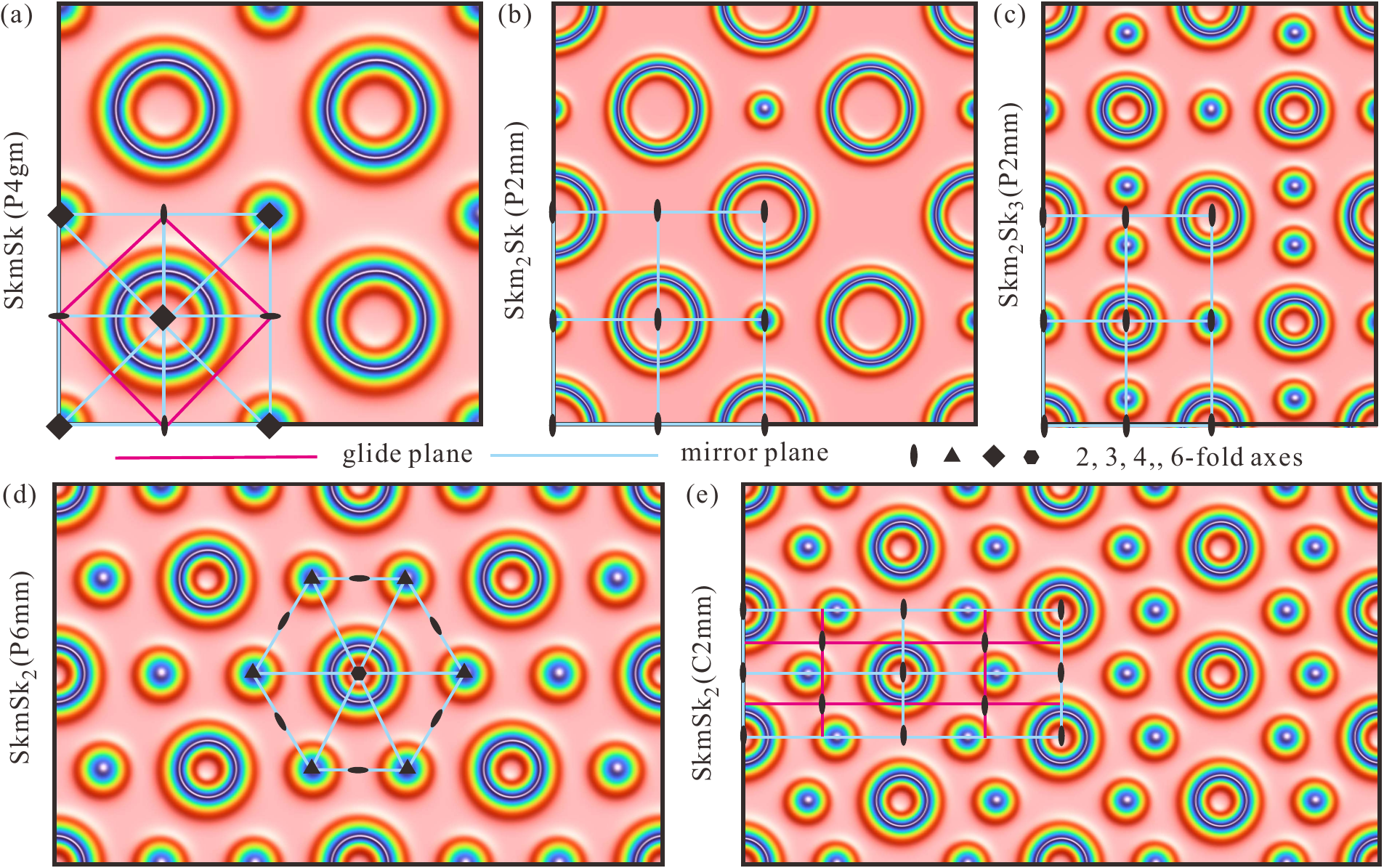}
  \caption{\label{fig04} 
  Examples of skyrmionium-based meta-matter with varying ratios of skyrmioniums and skyrmions. 
  (a) A staggered square lattice with an equal number of Skms and Sks (SkmSk). 
  (b) A square lattice Skm$_2$Sk with a reduced fraction of Sks. 
  (c) A compound Skm$_2$Sk$_3$ with nonuniform Skm sizes and slight elongation. 
  (d), (e) Polymorphs of the SkmSk$_2$ meta-matter exhibiting different planar symmetries. 
  The corresponding symmetry elements are indicated within one unit cell for each mixed lattice. 
  These structures exemplify the diversity of topological stoichiometries and plane-group symmetries possible in the skyrmionium–skyrmion compounds.
}
\end{figure*}

\begin{figure*}
  \centering
  \includegraphics[width=0.99\linewidth]{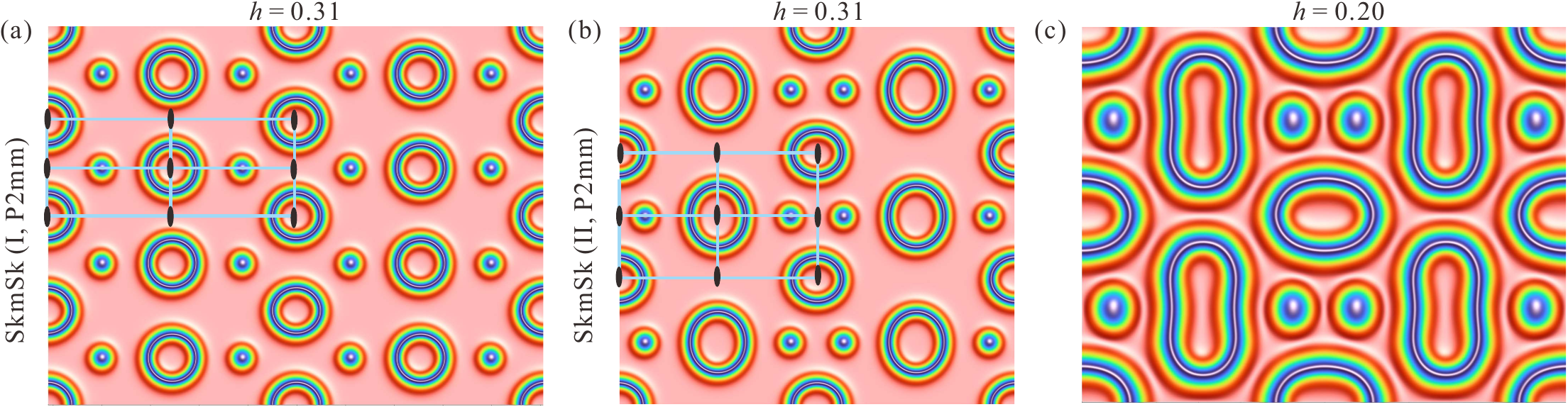}
  \caption{\label{fig04B} 
  Additional polymorphs of the SkmSk meta-matter composed of equal numbers of skyrmioniums and skyrmions . 
  (a) Configuration in which each skyrmion is surrounded by three skyrmioniums. 
  (b) Arrangement where two skyrmions are placed within a ``well'' enclosed by four skyrmioniums. 
  (c) A SkmSk polymorph stabilized by elongated skyrmioniums within the parameter region where the spiral state represents the local energy minimum. 
  These examples illustrate the structural versatility of the mixed Skm--Sk meta-matter and its ability to retain metastability even under conditions favoring other modulated phases.
  }
\end{figure*}

\subsection{Unstable skyrmion lattices}
In the same way as deduced for skyrmioniums, the hexagonal SkL and isolated skyrmions exist as detached branches of solutions on both sides of the critical anisotropy value $k_u(B)$ for $h=0$ [Fig.~\ref{fig03A}]. 

With decreasing $k_u$, the cores of isolated skyrmions expand [Fig.~\ref{fig03A}(c), (d)], and the localized skyrmions disappear as solutions before reaching a negative eigen-energy. Here, $\theta$ denotes the polar angle of the magnetization vector, and $r$ is the radial coordinate.
The characteristic core size $R_0$, defined according to the Lilley criterion [blue dashed lines in Fig.~\ref{fig03A}(a)], diverges at $k_u(B)$. 
Similarly, the hexagonal SkL expands [green solid lines in Fig.~\ref{fig03A}(a)] upon approaching the critical point from the left [Fig.~\ref{fig03A}(c)].

In the presence of an applied magnetic field $h>0$, however, isolated skyrmions can condense into a lattice as the uniaxial anisotropy constant $k_u$ decreases [Fig.~\ref{fig03A}(a)]. 
In this sense, the magnetic field $h>0$ bridges the two detached branches of skyrmion solutions. 
The solid red and green lines in Fig.~\ref{fig03A}(a) show the dependencies of the characteristic core size and SkL period on $k_u$, respectively. 
At the intersection point of the red and blue curves—corresponding to the condensation of isolated skyrmions into a lattice—the skyrmion core exhibits a sudden jump, while the equilibrium lattice period diverges.

As noted for the SkmL, the considered SkL exhibits an energy minimum only when the unit-cell dimensions are fixed, $p_x = \sqrt{3}\,p_y$ (or vice versa). However, the stability of the SkL must be examined with respect to arbitrary variations of the cell parameters. Figure~\ref{fig03B}(b) presents a characteristic energy “fingerprint,” similar to that shown in Fig.~\ref{fig02}. It is clearly seen that no local energy minimum corresponds to the hexagonal SkL, and the skyrmions elongate continuously into the spiral state [Fig. \ref{fig03B} (b)].

\begin{figure*}
  \centering
  \includegraphics[width=0.99\linewidth]{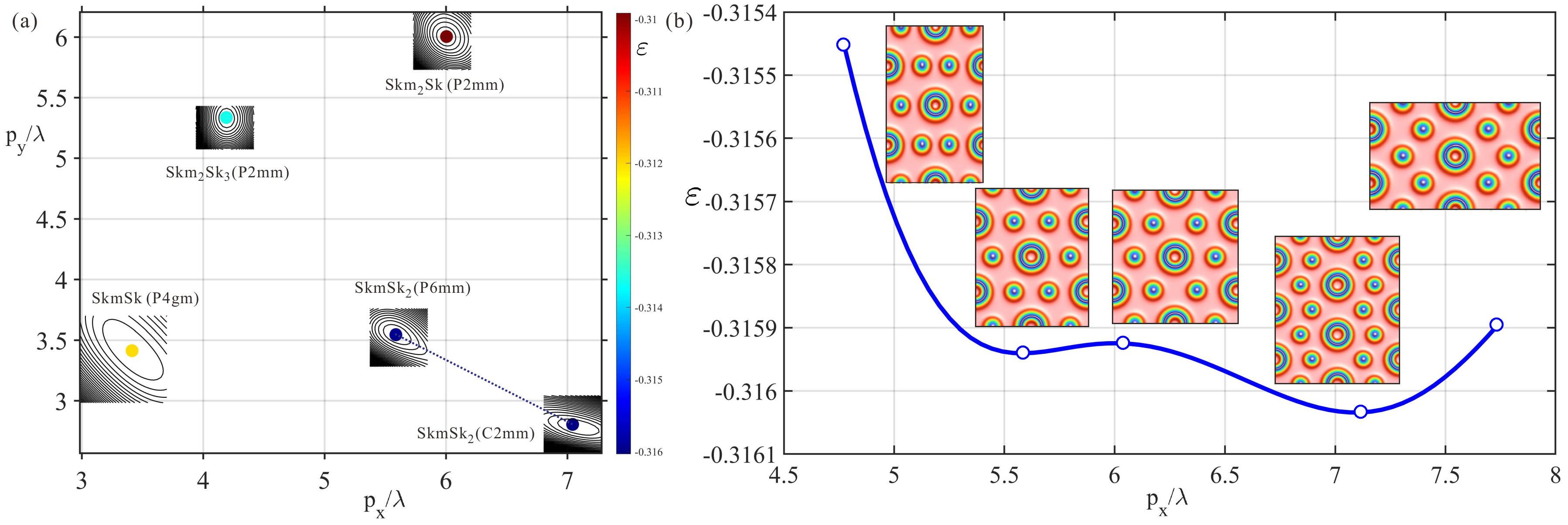}
  \caption{\label{fig05}
  (a) Equilibrium lattice parameters for various mixed Skm–Sk lattices. $h=0.31$, $k_u=0$.
  The color scale represents the energy density~$\varepsilon$, while the black contours surrounding each point serve as energy fingerprints, confirming that each configuration corresponds to a local energy minimum. 
  (b) Energy density profile along the transformation path connecting two polymorphs of the SkmSk$_2$ meta-matter. 
  The two states are separated by only a small energy barrier, indicating their close energetic competition and possible coexistence. 
  Representative spin configurations along the transformation path are shown as insets.
}
\end{figure*}

\section{Skyrmionium meta-matter: Mixed skyrmion-skyrmionium lattices}

The instability of pure skyrmionium lattices motivates the exploration of compound lattices in which skyrmioniums and ordinary skyrmions coexist. Mixed Skm–Sk arrangements combine elements with distinct topology ($Q \approx 0$ and $Q = -1$) and therefore offer the possibility that mutual interactions stabilize new periodic states inaccessible to pure Skm or Sk ensembles, thereby suppressing the elongation mode responsible for SkmL instability. Beyond their fundamental interest, such mixed lattices may preserve the favorable device-oriented properties of skyrmioniums (such as reduced transverse deflection) while inheriting the condensation-prone nature of conventional skyrmions, potentially yielding robust, tunable meta-matter for magnonics and spintronics.

We, however, refrain from detailed speculations on possible experimental routes for realizing such skyrmionium-based meta-matter. 
Briefly, one may envisage constructing the Skm--Sk meta-matter within the parameter-space region located below the line $D$--$B$ and above the line $c$--$B$ in the phase diagram [Fig.~\ref{fig01}(c)], where a moderate uniaxial anisotropy is essential for stability. 
Below the line $D$--$B$, the stable SkL phase tends to expand, keeping skyrmions at larger separations, and may thus serve as a convenient template for placing skyrmioniums in between skyrmions or for locally modifying nearly isolated skyrmions to transform them into skyrmioniums.
In layered systems with interfacially induced DMI, local writing of individual textures could be performed using a spin-polarized STM tip—via localized spin-polarized current and/or electric-field or thermal pulses—or through spin-current injection from nanocontacts. 
These techniques have been experimentally demonstrated to produce and control single skyrmions and, in suitably tailored current geometries, can also generate skyrmioniums. 
For related experimental demonstrations of local skyrmion writing, see Refs.~\onlinecite{Romming2013,Wieser2017}, and for theoretical proposals on skyrmionium nucleation, see Ref.~\onlinecite{Obadero2020}. 
In the following, we focus on well-balanced, theoretically engineered configurations corresponding to local energy minima.

 As a representative point on the phase diagram, we choose $(0, 0.31)$, which lies above the region of spiral stability and ensures that the quasi-atoms—skyrmioniums and skyrmions—retain their circular shape.

Figure~\ref{fig04} summarizes several Skm–Sk compounds with different topological stoichiometries (we use the term topological stoichiometry to denote the ratio of skyrmions to skyrmioniums within a composite lattice). The corresponding symmetry elements are also indicated, allowing the assignment of their plane groups. 

In the most obvious arrangement [Fig. \ref{fig04} (a)], skyrmioniums and ordinary skyrmions occupy alternating sites of a square unit cell, forming a regular checkerboard pattern. Each Skm is surrounded by four Sks on their nearest-neighbor positions, and vice versa. This configuration benefits from mutual stabilization: the surrounding Sks prevent the elongation of Skms, while the presence of Skms modifies the local interactions of Sks, potentially reducing their tendency to condense into a dense lattice. The checkerboard lattice represents a simple yet robust mixed Skm--Sk structure that preserves the quasi-circular shape of Skms and ensures a well-defined periodicity, making it an attractive candidate for experimental realization and magnonic applications.

Interestingly, even a small fraction of skyrmions is sufficient to ensure the stability of the compound, as exemplified by the Skm$_2$Sk structure with a square unit cell [Fig.~\ref{fig04}(b)]. 
In this configuration, two skyrmioniums share a single skyrmion within the unit cell. A skyrmion acts as a pin or “nail,” preventing the elongation of the surrounding Skms and preserving their circular shape. The skyrmion is strategically positioned to mediate interactions between neighboring Skms, thereby reinforcing the overall structural integrity. 
This, however, raises the question of the minimal fraction of Sks required to prevent the elongation of Skms and stabilize higher-order Skm$_n$Sk compounds with $n>2$. While the exact critical ratio depends on the lattice geometry and the underlying control parameters, preliminary simulations suggest that even a single skyrmion per several Skm units can act as an effective “pin,” suppressing the elongation mode (to be elaborated elsewhere).

Figs.~\ref{fig04} (c)--(e) illustrate Skm--Sk compounds with an increasing number of skyrmions. Notably, the Skm$_2$Sk$_3$ compound exhibits skyrmioniums of different sizes with slight elongation.

As a matter of fact, each topological stoichiometry class implies several possible realizations (polymorphs). For example, for SkmSk$_2$ compounds [Fig. \ref{fig04} (d), (e)], two planar tessellations are possible, motivated either by regular (hexagonal) space filling or by interactions among solitons of the same kind (striped / lamellar arrangement). Here, however, we focus on the configurations corresponding to the deepest energy minimum for each composition ratio.

In this sense, the spin configurations presented in Fig.~\ref{fig04} are by no means exhaustive. 
A broad variety of regular arrangements can be constructed by placing skyrmions into ``wells'' surrounded by three or four skyrmioniums, thereby generating numerous stoichiometric combinations. 
In the general case, domains with different local stoichiometries may emerge if skyrmions occupy their positions in a partially random manner. 
As an additional example, Fig.~\ref{fig04B} illustrates several polymorphs of the SkmSk meta-matter distinct from the square arrangement of constituents shown in Fig.~\ref{fig04}(a). 
Interestingly, some of these polymorphs preserve their metastability even below the line $c$--$B$ in the phase diagram [Fig.~\ref{fig01}(c)], where the spiral state constitutes a significantly deeper global energy minimum. 
While the square SkmSk lattice maintains nearly circular soliton profiles, the polymorph displayed in Fig.~\ref{fig04B}(c) represents a balanced configuration characterized by elongated skyrmioniums and slightly deformed skyrmions, highlighting the structural adaptability of the mixed Skm--Sk meta-matter.

Obviously, with an increasing number of skyrmions, the energy of a compound decreases and approaches that of the skyrmion lattice, since at these control parameters the SkL represents the global minimum of the system. Fig.~\ref{fig05} (a) shows the equilibrium parameters $p_x/\lambda$ and $p_y/\lambda$ of unit cells for different stoichiometries. The color of the points indicates the energy density, whereas the surrounding ``fingerprint'' denotes that these are local energy minima.

We can also controllably transform between different polymorphs of SkmSk$_2$ by tuning the lattice spacings $\Delta_x$ and $\Delta_y$. 
Fig.~\ref{fig05}(b) shows the energy density along the trajectory indicated by the dotted blue line in Fig.~\ref{fig05}(a), together with snapshots of the spin configurations at selected points along this path.
In experiments, such transformations between different polymorphs with the same topological stoichiometry could potentially be achieved by tuning the effective lattice spacings. For instance, applying uniaxial or hydrostatic pressure, or engineering confinement in nanostructures, may effectively compress or stretch the lattice, thereby favoring one polymorph over another. Alternatively, local modifications of the magnetic anisotropy or the Dzyaloshinskii-Moriya interaction could also induce similar structural rearrangements.

\begin{figure*}
  \centering
  \includegraphics[width=0.99\linewidth]{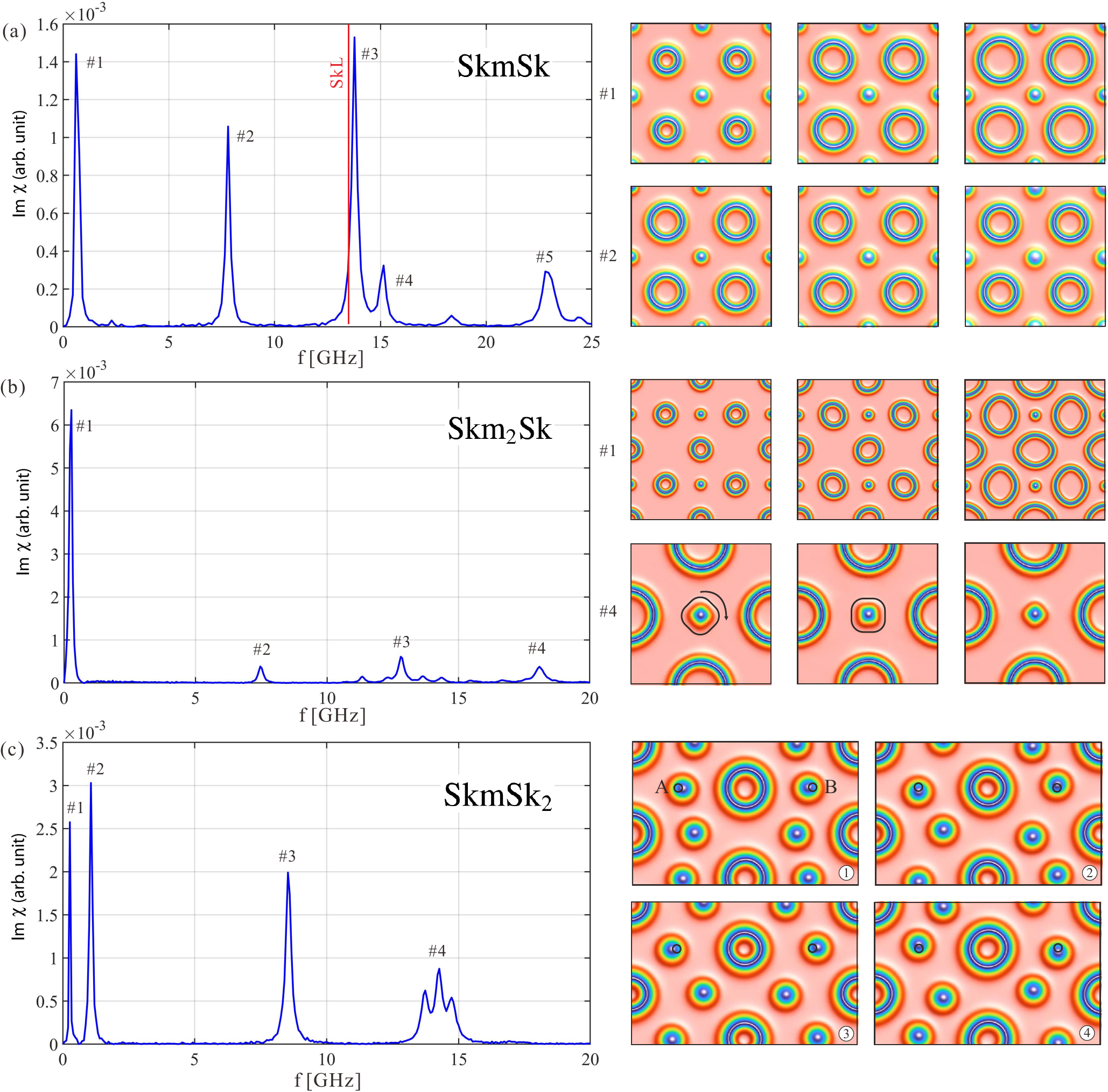}
  \caption{\label{fig06} 
  Imaginary parts of the out-of-plane dynamical susceptibility, $\mathrm{Im}\,\varkappa(\omega)$, for mixed Skm–Sk lattices with different topological stoichiometries: (a) SkmSk crystals; (b) Skm$_2$Sk crystals; (c) SkmSk$_2$ crystals. 
  Each spectrum exhibits multiple resonance modes that, in general, involve not only the breathing oscillations of skyrmions and skyrmioniums, but also shape deformations of skyrmions, their lateral displacements, and circular (orbital) motions within confining potential wells (see text for details). 
  The snapshots in the right column illustrate representative spin configurations at selected time moments for chosen modes, whereas the Supplementary Videos provide a more complete visualization of the observed dynamical behaviors.
  }
\end{figure*}

\section{Collective spin dynamics of skyrmionium meta-matter}

In this section, we examine the collective spin dynamics of the previously described skyrmionium-based meta-matter for both in-plane and out-of-plane orientations of the applied ac magnetic field.
We closely follow the numerical procedure described in Ref.~\onlinecite{Mochizuki}; therefore, detailed methodological explanations are omitted here for conciseness.

All simulations were performed using \textsc{mumax3} with the material parameters corresponding to Co/Pt multilayers, as introduced earlier in Sect. \ref{sect:model}.  
To excite the system, we apply short in-plane and out-of-plane $\delta$-function  pulses of the magnetic field $\delta(t)\mathbf{H}^\omega$ at $t=0$ and subsequently track the time evolution of the magnetization in various mixed Skm--Sk crystals.  
The absorption spectrum, represented by the imaginary part of the dynamical susceptibility $\mathrm{Im}\,\varkappa(\omega)$, is obtained via the Fourier transform of the spatially averaged magnetization components.  
A relatively small damping parameter, $\alpha = 0.01$, is employed to ensure well-resolved resonance peaks in $\mathrm{Im}\,\varkappa(\omega)$ and to accurately capture the intrinsic dynamical modes of the system.

\subsection{Breathing modes of the skyrmionium meta-matter}

The calculated spectra for $\mathbf{H}^\omega$ applied parallel to the $z$ axis are summarized in Fig.~\ref{fig06}. 
Whereas the ordinary SkL exhibits only a single breathing mode, the skyrmionium meta-matter typically displays several resonance peaks of comparable intensity.

We begin with an exhaustive analysis of the breathing modes of the staggered Skm--Sk crystal [Fig. \ref{fig06} (a)]; all corresponding videos are provided in the Supplementary Materials.

The first mode \#1 in the spectrum ($f = 0.612$~GHz) corresponds to the breathing oscillations of the skyrmioniums. Remarkably, the frequency of the ac magnetic field required to excite this mode is significantly lower than that for the pure SkL (indicated by the red line, $f = 13.5$~GHz). Skyrmions, in contrast, are not excited in the first mode and remain essentially unaffected [snapshots in Fig. \ref{fig06} (a)].

The second mode \#2, observed at $f = 7.8$~GHz, represents the opposite situation, with skyrmions undergoing breathing oscillations while the skyrmioniums remain inert [snapshots in Fig. \ref{fig06} (a)].

The third ($f=13.77$GHz) and fourth modes ($f=15.15$GHz) correspond to the simultaneous breathing of both skyrmions and skyrmioniums, occurring either in-phase or in anti-phase (see Supplementary Videos for details). Remarkably, we also identify a fifth mode at a much higher frequency, $f = 22.8~\mathrm{GHz}$.
This mode represents a hybrid breathing–rotational excitation, in which skyrmioniums undergo radial breathing while skyrmions deform into a square-like shape and perform rotational motion within their potential wells \cite{Kravchuk2018}. 

\begin{figure}[t]
  \centering
  \includegraphics[width=0.999\linewidth]{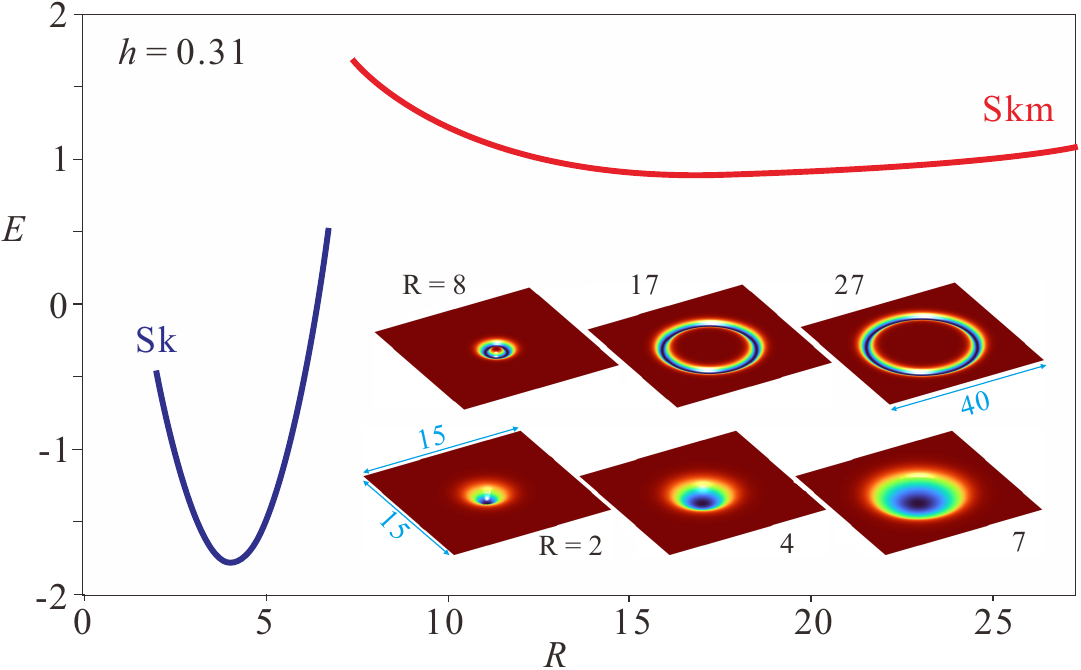}
  \caption{\label{fig08}
  Eigen-energies of isolated skyrmioniums (red) and skyrmions (blue) as functions of the radius $R$ at which the magnetization is pinned. For both solitons  the magnetization was fixed at $m_z = 0$ along a circle of radius $R$. 
  Insets show representative soliton profiles for different radii. $h=0.31, k_u=0$.}
\end{figure}

Note that the characteristic exchange frequency, estimated from 
$f_\mathrm{ex} = (\gamma / 2\pi)\mu_0 H_\mathrm{ex} \approx 58~\mathrm{GHz}$ 
with $H_\mathrm{ex} \sim 2A / (\mu_0 M_s L_D^2) \approx 1.6\times10^6~\mathrm{A/m}$ 
and $\mu_0 H_\mathrm{ex} \approx 2.1~\mathrm{T}$, corresponds to the local ferromagnetic resonance scale of a homogeneous magnetization. The collective breathing and hybrid modes of the skyrmionium meta-matter,
being governed by weaker restoring fields, are therefore expected to appear at considerably lower frequencies, typically in the range of a few to several tens of~GHz.

If we decrease the skyrmion fraction and examine the Skm$_2$Sk crystals, we still identify the first breathing mode of skyrmioniums at $f = 0.28~\mathrm{GHz}$ [Fig.~\ref{fig06}(b)]. 
During this breathing motion, individual skyrmioniums become elongated as they attempt to fill the voids left by the absence of skyrmions [see snapshots in Fig.~\ref{fig06}(b)]. 
The second mode, appearing at $f = 7.47~\mathrm{GHz}$, again corresponds to the breathing of skyrmions while skyrmioniums remain largely unaffected, although its intensity is considerably lower. 
The third mode, at $f = 12.8~\mathrm{GHz}$, corresponds to an in-phase breathing oscillation of skyrmions and skyrmioniums. In contrast, the fourth mode, observed at $f = 18.08~\mathrm{GHz}$, features the rotation of deformed, square-like skyrmions driven by the breathing motion of surrounding skyrmioniums, as previously discussed for Skm–Sk crystals [see snapshots in Fig.~\ref{fig06}(b) and Supplementary Videos]. In the present case as well, the breathing frequency of the pure SkL roughly coincides with the mode in which both types of solitons undergo synchronous breathing.

The most intricate modes, however, are found for an increased fraction of skyrmions. For the SkmSk$_2$ (C2mm) crystal, we first identify the modes described previously: 
mode~$\#2$ corresponds to the breathing oscillation of skyrmioniums ($f=1.07$~GHz); 
mode~$\#3$ represents the breathing of skyrmions ($f=8.53$~GHz); 
and mode~$\#4$ ($f=14.27$GHz) corresponds to the simultaneous breathing of both skyrmions and skyrmioniums. 
The latter mode is split into three closely spaced peaks, accompanied by a slight elongation of the skyrmioniums and a rotation of the skyrmion centers (see Supplementary Videos).  

The most unexpected and intriguing feature is the low-frequency mode~$\#1$ ($f=0.27$GHz). 
This mode can be described as a mixed excitation where skyrmioniums undergo a breathing motion (changing their size), while the skyrmions do not vary in size but perform circular orbital motion within their potential wells [see Supplementary Videos]. Snapshots $1-4$ in Fig.~\ref{fig06}(c) show different time moments of such a circular skyrmion motion. The trajectory spanned by the skyrmion centers is shown by black circles for two exemplary skyrmions A and B, which oscillate in anti-phase, i.e. when the skyrmion A is to the right from the trajectory center, the skyrmion B is to the left. In this sense, the breathing skyrmioniums pump such a skyrmion movement. Thus, this mode can be classified as  a coupled breathing–orbiting excitation, where skyrmioniums and skyrmions exhibit dynamically intertwined radial and orbital motions.

A conceptually similar mode is also observed for the Skm$_2$Sk$_3$ crystal. 
While solitary skyrmions remain intact within their wells, pairs of skyrmions confined in the same well perform circular orbital motion, driven by the periodic inflation and deflation of the surrounding skyrmioniums [see Supplementary Videos].

To summarize, in the Skm–Sk crystals, a rich hierarchy of collective excitation modes emerges as a result of the complex internal structure of the composite lattice and the mutual dynamical coupling between its constituents. The simplest modes correspond to \emph{pure breathing oscillations}, where skyrmioniums and/or skyrmions expand and contract radially while preserving their circular symmetry. At higher frequencies, \emph{shape-deformation modes} appear, characterized by transient distortions of the soliton profiles 
reflecting the hybridization of radial and azimuthal spin-wave components. In addition, the system hosts \emph{translational and rotational modes} in which the individual skyrmions, without significantly changing their size, undergo lateral displacements or orbital rotations within their confining potential wells, often driven or ``pumped'' by the breathing motion of neighboring skyrmioniums. Together, these distinct dynamical regimes underline the strongly coupled nature of the Skm–Sk lattice and the multiplicity of internal degrees of freedom accessible in such chiral meta-matter.

In addition, the pronounced difference between the breathing-mode frequencies of skyrmions and skyrmioniums can be understood in terms of the curvature of their respective energy landscapes. 
To quantify this effect, we computed the total energy of isolated solitons as a function of their enforced size by pinning the magnetization along a circular boundary at $m_z = 0$ [Fig.~\ref{fig08}]. 
The resulting $E(R)$ dependencies, where $E$ denotes the eigen-energy of each soliton, reveal that the skyrmionium energy landscape is remarkably shallow, whereas that of the skyrmion is considerably steeper. 
This indicates a much smaller effective stiffness, $\partial^2 E / \partial R^2$, for the skyrmionium, resulting in a lower characteristic frequency of its breathing mode. 
Conversely, the steep confinement potential of the skyrmion leads to a stronger restoring force and thus a higher breathing frequency. 
This correlation between the curvature of the energy profile and the oscillation frequency follows the harmonic-oscillator analogy, $\omega_{\mathrm{breath}} \propto \sqrt{k / M_{\mathrm{eff}}}$, providing a natural explanation for the observed mode hierarchy. 
Moreover, the energy curves suggest a topological connection between the two solitons: an expanding skyrmion may lower its energy by nucleating a nested core of opposite polarity, thereby transforming into a skyrmionium; conversely, a contracting skyrmionium may collapse into a single skyrmion.

\begin{figure}
  \centering
  \includegraphics[width=0.999\linewidth]{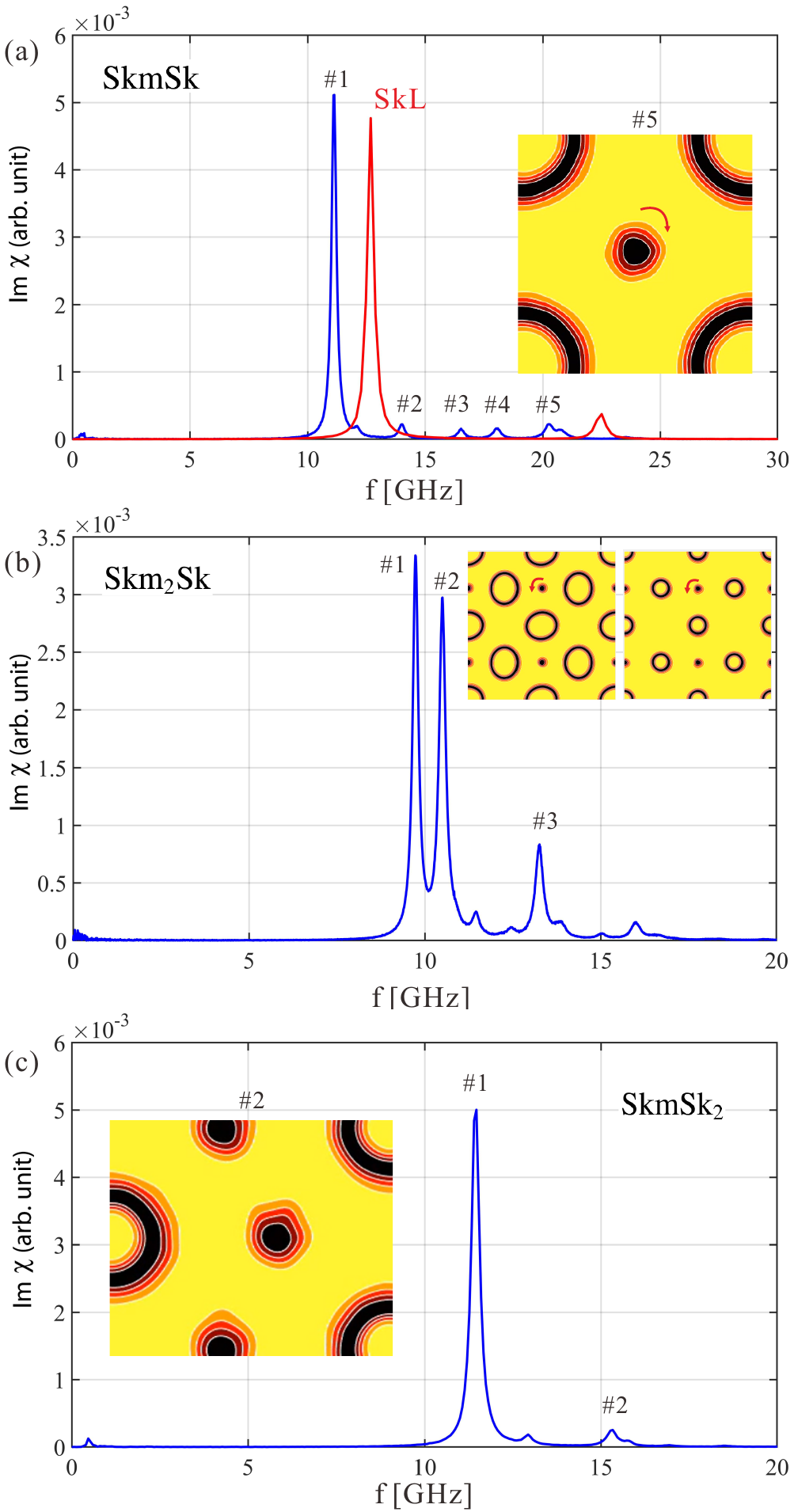}
  \caption{\label{fig07} 
  Imaginary parts of the in-plane dynamical susceptibility, $\mathrm{Im}\,\varkappa(\omega)$, for mixed Skm–Sk lattices with different topological stoichiometries: 
  (a) SkmSk crystal; (b) Skm$_2$Sk crystal; (c) SkmSk$_2$ crystal. 
  Each spectrum displays several resonance peaks corresponding to collective excitation modes that combine the orbital motion of skyrmions—characteristic of pure SkLs—with the breathing dynamics of skyrmioniums. 
  The coupling between these two degrees of freedom gives rise to hybridized magnonic modes whose structure and frequency strongly depend on the Skm-to-Sk ratio. 
  Representative spin configurations of selected modes are shown as snapshots, while the Supplementary Videos provide a more comprehensive visualization of the associated dynamical patterns.
  }
\end{figure}

\subsection{Collective modes activated by the in-plane ac magnetic field}

The calculated spectra for $\mathbf{H}^\omega$ applied parallel to the $x$ axis are summarized in Fig.~\ref{fig07}. 
While the hexagonal SkL is known to exhibit two distinct resonances in the in-plane dynamical susceptibility---one corresponding to the counterclockwise (CCW) and the other to the clockwise (CW) rotation of the skyrmion cores [the spectrum is shown by the red curve in Fig.~\ref{fig07}(a)]~\cite{Mochizuki}---the spectra of the skyrmionium-based meta-matter reveal a far richer structure, featuring multiple resonance peaks that reflect its composite nature (blue curves). 
To identify the origin of each spin-wave mode, we excite the system with a weak oscillating magnetic field at the corresponding resonant frequency and analyze the resulting spin dynamics.

As before, we start with the SkmSk crystal. 
In the first mode ($f = 11.12~\mathrm{GHz}$), both Skm and Sk undergo a CCW rotation, resembling the first mode of the pure SkL observed at $f = 12.68~\mathrm{GHz}$. 
Here, it is important to underline the \textit{in-phase} rotation of skyrmions and the outer Skm rings (both carrying $Q = -1$), in contrast to the \textit{anti-phase} rotation of the skyrmions and the skyrmionium cores ($Q = +1$). 
The second mode ($f = 14~\mathrm{GHz}$) corresponds to the clockwise rotation of both Skm and Sk, thus resembling the second SkL mode found at $f = 22.5~\mathrm{GHz}$. 
The third mode ($f = 16.53~\mathrm{GHz}$) also exhibits CCW rotation; however, in this case, the skyrmions and the Skm cores rotate in-phase, whereas the Skm rings oscillate in anti-phase.
In the fourth mode ($f = 18.1~\mathrm{GHz}$), the skyrmions rotate counterclockwise, the Skm cores also rotate CCW but in anti-phase, whereas the Skm rings rotate CCW in-phase with the skyrmions. 
The fifth mode ($f = 20.25~\mathrm{GHz}$) is characterized by a pronounced deformation of the skyrmions—their cores acquire a triangular shape while their outer boundaries become hexagonal [see inset in Fig. \ref{fig07}(a)]—and by a rotational motion around their centers, driven by the slight rotational oscillations of the surrounding skyrmioniums.

A reduced fraction of skyrmions in the Skm$_2$Sk crystal leads to several notable modifications of the excitation spectrum. 
While the first mode (\#1, $f = 9.73~\mathrm{GHz}$) qualitatively resembles the lowest-frequency mode of the SkmSk counterpart, the second mode (\#2, $f = 10.5~\mathrm{GHz}$) represents a pronounced deviation. 
Here, the skyrmions and the Skm rings undergo a counterclockwise in-phase rotational motion, which is dynamically coupled to the breathing oscillations of the skyrmioniums [see snapshots in Fig.~\ref{fig07} and Supplementary Videos]. 
The third mode (\#3, $f = 13.26~\mathrm{GHz}$) reveals an even more intricate hybrid character: in addition to the breathing motion of skyrmioniums, it exhibits a reversal of the rotational sense of the skyrmions, indicating a transition between distinct dynamical regimes.

In contrast, an increased fraction of skyrmions leads to a pronounced simplification of the excitation spectrum. 
The first mode ($f = 11.45~\mathrm{GHz}$) of the SkmSk$_2$ crystal corresponds to the canonical CCW rotational motion of both skyrmioniums and skyrmions, similar to the fundamental CCW mode of the pure SkL. 
The second mode ($f = 15.3~\mathrm{GHz}$) exhibits a more complex mixed character: the skyrmioniums oscillate back and forth along the direction of the ac magnetic field, while the skyrmions—deformed into slightly pentagonal shapes due to the local field gradients—undergo rotational motion around their centers  [see inset in Fig. \ref{fig07} (c)]. 
Such deformation-assisted rotation highlights the interplay between translational and internal degrees of freedom in densely packed Skm--Sk lattices.

\section{Conclusions}

In this work, we have systematically investigated the stability and collective dynamics of skyrmionium-based meta-matter within the minimal chiral ferromagnetic model incorporating exchange, Dzyaloshinskii–Moriya interaction, uniaxial anisotropy, and an external magnetic field. Our analysis reveals several fundamental aspects governing the existence and transformation of topological magnetic lattices.

First, we demonstrated that a pure skyrmionium lattice  is intrinsically unstable, independent of its initial symmetry—whether hexagonal or square. In all cases, the SkmL configuration spontaneously relaxes into the one-dimensional modulated spiral state, which represents the true thermodynamic ground state within the corresponding region of the $(h, k_u)$ parameter space. The boundary of this instability region can be indirectly traced through the behavior of isolated skyrmioniums: as their eigen-energy approaches zero, the structures expand indefinitely without ever crossing into a negative mode that would allow condensation into a stable SkmL. A similar instability mechanism operates in the hexagonal skyrmion lattice, which becomes unstable at zero field and continuously deforms into the spiral state. 

Despite the inherent instability of pure SkmL phase, the skyrmionium-based meta-matter can attain metastable configurations when skyrmioniums intermix with skyrmions, forming composite crystalline states Skm$_n$Sk$_m$. Among them, the staggered square lattice (SkmSk) with plane group $P4gm$ emerges as a particularly representative example. Further exploration of such mixed lattices reveals a family of polymorphs with varying ratios of skyrmions and skyrmioniums (e.g., Skm$_2$Sk, Skm$_2$Sk$_3$, SkmSk$_2$), each corresponding to distinct local minima of the magnetic energy functional. The coexistence and near-degeneracy of these polymorphs highlight the remarkable configurational diversity and metastability inherent to skyrmionium meta-matter.

Finally, we analyzed the low-frequency spin excitation spectra of these composite states. The mixed Skm–Sk lattices support a rich hierarchy of collective modes that combine internal oscillations of individual constituents with long-range coupling across the lattice. In particular, several modes share a similar rotational character of the Sk and Skm units but differ in their collective phase relationships, reflecting the underlying lattice symmetry and coupling topology. These complex excitations underscore the potential of skyrmionium-based meta-matter as a versatile platform for magnonic applications, where tunable coupling between topological quasiparticles enables the engineering of multifunctional spin-wave responses.

In summary, our findings establish the instability of pure SkmL and SkL phases, reveal the emergence of a broad family of metastable mixed lattices, and elucidate their distinctive collective dynamics. Together, these results provide a coherent framework for understanding and controlling skyrmionium-based magnetic meta-matter and its potential use in reconfigurable spintronic and magnonic devices.


\begin{thebibliography}{99}

\bibitem{manton_sutcliffe}
N. Manton and P. Sutcliffe, \textit{Topological Solitons} (Cambridge University Press, Cambridge, 2004).

\bibitem{shnir}
Y. M. Shnir, \textit{Topological and Non-topological Solitons in Scalar Field Theories} (Cambridge University Press, Cambridge, 2018).

\bibitem{Volovik}
G. E. Volovik and V. P. Mineev, \textit{Particle-like solitons in superfluid $^3$He phase}, Sov. Phys. JETP \textbf{46}, 401 (1977).

\bibitem{solitons}
R. Rajaraman, \textit{Solitons and Instantons: An Introduction to Solitons and Instantons in Quantum Field Theory} (North-Holland, Amsterdam, 1982).

\bibitem{JETP89}   A. N. Bogdanov and D. A. Yablonsky, \textit{Thermodynamically stable vortices in magnetically ordered crystals. Mixed state of magnetics}, 
 Zh. Eksp. Teor. Fiz. {\textbf 95}, 178 (1989) [Sov. Phys. JETP \textbf{68}, 101 (1989)].

\bibitem{Bogdanov94} A. Bogdanov and A. Hubert, \textit{Thermodynamically stable magnetic vortex states in magnetic crystals}, J. Magn. Magn. Mater. \textbf{138}, 255 (1994).

\bibitem{NT} N. Nagaosa and Y. Tokura, \textit{Topological properties and dynamics of magnetic skyrmions},  Nat. Nanotechnol. \textbf{8}, 899 (2013).


\bibitem{Faddeev}
L. Faddeev and A. Niemi, \textit{Stable knot-like structures in classical field theory}, Nature \textbf{387}, 58 (1997).

\bibitem{Bott}
R. Bott and L. W. Tu, \textit{Differential Forms in Algebraic Topology} (Springer, New York, 1995).

\bibitem{Kovalev2018} Kovalev, A. A., Sandhoefner, S. \textit{Skyrmions and Antiskyrmions in Quasi-Two-Dimensional Magnets}. Front. Phys. \textbf{6}, \textrm{98} (2018).

\bibitem{Bogdanov99} A. Bogdanov and A. Hubert,  \textit{The stability of vortex-like structures in uniaxial ferromagnets},  J. Magn. Magn. Mater. \textbf{195}, 182 (1999).

\bibitem{Dz58} I. E. Dzyaloshinskii, \textit{A thermodynamic theory of weak ferromagnetism of antiferromagnetics}, J. Phys. Chem. Sol. \textbf{4}, 241 (1958).

\bibitem{Moriya} T. Moriya, \textit{Anisotropic Superexchange Interaction and Weak Ferromagnetism}, Phys. Rev. \textbf{120}, 91 (1960).

\bibitem{Wiesendanger2016} R. Wiesendanger, \textit{Nanoscale magnetic skyrmions in metallic films and multilayers: A new twist for spintronics}, Nat. Rev.
Mater. \textbf{1}, 16044 (2016).

\bibitem{Muehlbauer09} S. M\"uhlbauer, B. Binz, F. Jonietz, C. Pfleiderer, A. Rosch, A. Neubauer, R. Georgii, P. B\"oni, 
\textit{Skyrmion lattice in a chiral magnet}, 
Science \textbf{323}, 915 (2009).

\bibitem{FeGe} H. Wilhelm, M. Baenitz, M. Schmidt, U. K. Roessler, A. A. Leonov, A. N. Bogdanov,  \textit{Precursor phenomena at the magnetic ordering of the cubic helimagnet FeGe}, Phys. Rev. Lett. \textbf{107}, 127203 (2011).

\bibitem{Damien} D. McGrouther, R. J. Lamb, M. Krajnak, S. McFadzean, S. McVitie, R. L. Stamps, A. O. Leonov, A. N. Bogdanov, and Y. Togawa, 
\textit{Internal structure of hexagonal skyrmion lattices in cubic helimagnets}, New J. of Phys. \textbf{18}, 095004 (2016).

\bibitem{Birch:2020} M. T Birch, D. Cort{\'e}s-Ortu{\~n}o,  L. A. Turnbull, M. N. Wilson, F. Gro{\ss}, N. Tr{\"a}ger,  A. Laurenson, N. Bukin, S. H. Moody, M. Weigand, G. Sch{\"u}tz, H. Popescu, R. Fan, P. Steadman, J. A. T. Verezhak, G. Balakrishnan , J. C. Loudon, A. C. Twitchett-Harrison, O. Hovorka, H. Fangohr, F. Y. Ogrin, J. Gr{\"a}fe, and P. D. Hatton, 
\textit{Real-space imaging of confined magnetic skyrmion tubes}, Nat. Commun., \textbf{11},1726 (2020).

\bibitem{Romming2013} N. Romming, C. Hanneken, M.  Menzel, J. E. Bickel, B. Wolter, K. von Bergmann, A.  Kubetzka, and R. Wiesendanger, 
\textit{Writing and Deleting Single Magnetic Skyrmions}, 
Science \textbf{341}, 636 (2013).

\bibitem{Sampaio13} J. Sampaio, V. Cros, S. Rohart, A. Thiaville,  and A. Fert, 
\textit{Nucleation, stability and current-induced motion of isolated magnetic skyrmions in nanostructures}, 
Nat. Nanotechnol. \textbf{8}, 839844 (2013).

\bibitem{Tomasello14} E. M. R. Tomasello, R. Zivieri, L. Torres, M. Carpentieri, and G. Finocchio, 
\textit{A strategy for the design of skyrmion racetrack memories}, 
Sci. Rep. \textbf{4}, 6784 (2014).

\bibitem{Shigenaga} T. Shigenaga and A. O. Leonov, \textit{Harnessing Skyrmion Hall Effect by Thickness Gradients in Wedge-Shaped Samples of Cubic Helimagnets}, Nanomaterials \textbf{13 (14)}, 2073 (2023).

\bibitem{Cortes-Ortuno} D. Cortes-Ortuno, W. Wang, M. Beg, R. A. Pepper, M.-A. Bisotti, R. Carey, M. Vousden, T. Kluyver, O. Hovorka, and H. Fangohr, 
\textit{Thermal stability and topological protection of skyrmions in nanotracks}, Sci. Rep. \textbf{7}, 1 (2017).

\bibitem{Schulz} T. Schulz, R. Ritz, A. Bauer, M. Halder, M. Wagner, C. Franz, C. Pfleiderer, K. Everschor, M.  Garst, A. Rosch, 
\textit{Emergent electrodynamics of skyrmions in a chiral magnet}, Nat. Phys. \textbf{8}, 301–304 (2012).

\bibitem{Jonientz} F. Jonietz, S. M\"uhlbauer, C. Pfleiderer, A. Neubauer, W. M\"unzer, A. Bauer, T. Adams, R. Georgii, P. B\"oni, R. A. Duine, K. Everschor, M. Garst, and A. Rosch, 
\textit{Spin Transfer Torques in MnSi at Ultralow Current Densities}, Science \textbf{330}, 1648–1651 (2010).

\bibitem{Wang16} W. Kang, Y. Huang, C. Zheng, W. Lv, Na Lei, Y. Zhang, X. Zhang, Y. Zhou, and W. Zhao, 
\textit{Voltage Controlled Magnetic Skyrmion Motion for Racetrack Memory},   Sci. Rep. \textbf{6}, 23164 (2016).
 
\bibitem{Fert2013} A. Fert, V. Cros, J. Sampaio, 
\textit{Skyrmions on the track}, Nat. Nanotechnol.  \textbf{8},  152 (2013).


\bibitem{Toscano} D. Toscano, J. Mendonca, A. Miranda, C. de Araujo, F. Sato, P. Coura, and S. Leonel, 
\textit{Suppression of the skyrmion Hall effect in planar nanomagnets by the magnetic properties engineering: Skyrmion transport on nanotracks with magnetic strips}, J. of Magn. and Magn. Mater. \textbf{504}, 166655 (2020).

\bibitem{Gobel} B. Gobel, A. Mook, J. Henk, and I. Mertig, 
\textit{Overcoming the speed limit in skyrmion racetrack devices by suppressing the skyrmion Hall effect}, Phys. Rev. B \textbf{99}, 020405(R) (2019).

\bibitem{deGennes} P. G. de Gennes, in \textit{Fluctuations, Instabilities, and Phase transitions}, edited by T. Riste, NATO ASI Ser. B, Vol. 2 (Plenum,
New York, 1975).

\bibitem{Mochizuki} M. Mochizuki,  \textit{Spin-Wave Modes and Their Intense Excitation Effects in Skyrmion Crystals}. \textrm{Phys. Rev. Lett.}  \textbf{108}, 017601 (2012). 

\bibitem{Desplat} L. Desplat and B. Dupe, \textit{Eigenmodes of magnetic skyrmion lattices}, Phys. Rev. B \textbf{107}, 144415 (2023). 



\bibitem{Leonov14} A. O. Leonov, U. K. Roessler, and M. Mostovoy, \textit{Target-skyrmions and skyrmion clusters in nanowires of chiral magnets}, EPJ Web Conf. \textbf{75}, 05002 (2014).

\bibitem{Komineas} S. Komineas, N. Papanicolaou, \textit{Skyrmion dynamics in chiral ferromagnets}, Phys. Rev. B \textbf{92}, 064412 (2015).

\bibitem{Nakamura} K. Nakamura and A. O. Leonov, \textit{ Mechanism of skyrmionium stability in quasi-two-dimensional chiral magnets }, Phys. Rev. B \textbf{110},094403  (2024). 

\bibitem{Kolesnikov18} A. G. Kolesnikov, M. E. Stebliy, A. S. Samardak and A. V. Ognev, 
\textit{Skyrmionium  high velocity without the skyrmion Hall effect}, Sci. Rep. \textbf{8}, 16966 (2018).

\bibitem{Wang20} J. Wang, J. Xia, X. Zhang, X. Zheng, G. Li, L. Chen, Y. Zhou, J. Wu, H. Yin, R. Chantrell, Y. Xu, 
\textit{Magnetic skyrmionium diode with a magnetic anisotropy voltage gating}, Appl. Phys. Lett. \textbf{117}, 202401 (2020).

\bibitem{Higgins} J. M. Higgins, R. Ding, J. P. DeGrave, and S. Jin, 
\textit{Signature of Helimagnetic Ordering in Single-Crystal MnSi Nanowires}, Nano Lett. \textbf{10}, 1605 (2010).

\bibitem{Butenko} A. B. Butenko, A. A. Leonov, A. N. Bogdanov, and U. K. Roessler, \textit{Theory of vortex states in magnetic nanodisks with induced Dzyaloshinskii-Moriya interactions}, Phys. Rev. B \textbf{80}, 134410 (2009).

\bibitem{Ponsudana21} M. Ponsudana, R. Amuda, R. Madhumathi, A. Brinda, N. Kanimozhi, 
\textit{Confinement of stable skyrmionium and skyrmion state in ultrathin nanoring}, Phys. B: Condens. Matter \textbf{618}, 413144 (2021).

\bibitem{Kent} N. Kent, R. Streubel, Charles-Henri Lambert, A. Ceballos, S.-G. Je, S. Dhuey, Mi-Young Im, F. B\"uttner, F. Hellman, S. Salahuddin, and P. Fischer, 
\textit{Generation and stability of structurally imprinted target skyrmions in magnetic multilayers}, Appl. Phys. Lett. \textbf{115}, 112404 (2019).

\bibitem{Zheng} F. Zheng, H. Li, S. Wang, D. Song, C. Jin, W. Wei, A. Kovacs, J. Zang, M. Tian, Y. Zhang, H. Du, and R. E. Dunin-Borkowski, 
\textit{Direct Imaging of a Zero-Field Target Skyrmion and Its Polarity Switch in a Chiral Magnetic Nanodisk}, Phys. Rev. Lett. \textbf{119}, 197205 (2017).

\bibitem{Zhang} S. Zhang, F. Kronast, G. van der Laan, and T. Hesjedal, 
\textit{Real-Space Observation of Skyrmionium in a Ferromagnet-Magnetic Topological Insulator Heterostructure}, Nano Lett. \textbf{18}, 1057 (2018).

\bibitem{Yang23} S. Yang, X. Li, Y. Zhao, Kai Wu, Z. Chu, J. Akerman, X. Xu, and Y. Zhou, 
\textit{Reversible conversion between skyrmions and skyrmioniums}, Nat. Commun. \textbf{14}, 3406 (2023).

\bibitem{Pwoalla23} L. Powalla, M. T. Birch, K. Litzius, S. Wintz, F. S. Yasin, L. A. Turnbull, F. Schulz, D. A. Mayoh, G. Balakrishnan, M. Weigand, X. Yu, K. Kern, G. Sch\"utz, and M. Burghard, 
\textit{Seeding and Emergence of Composite Skyrmions in a van der Waals Magnet}, Adv. Mater. \textbf{35}, 2208930 (2023).

\bibitem{Hagemeister} J. Hagemeister, A. Siemens, L. Rozsa, E. Vedmedenko, and R. Wiesendanger, 
\textit{Controlled creation and stability of $k\pi$ skyrmions on a discrete lattice}, Phys. Rev B \textbf{97}, 174436 (2018).

\bibitem{Jiang24} A. Jiang, Y. Zhou, X. Zhang, and M. Mochizuki, 
\textit{Transformation of a skyrmionium to a skyrmion through the thermal annihilation of the inner skyrmion}, Phys. Rev. Res. \textbf{6}, 013229 (2024).

\bibitem{Seki2012} S. Seki,  X. Z. Yu, S. Ishiwata, and Y. Tokura, 
\textit{Observation of Skyrmions in a Multiferroic Material}, 
Science \textbf{336}, 198 (2012).

\bibitem{Crisanti} M. Crisanti, A. O.  Leonov, R. Cubitt, A.  Labh, H. Wilhelm, M. P. Schmidt, C. Pappas,. \textit{Tilted spirals and low-temperature skyrmions in Cu$_2$OSeO$_3$}. \textrm{Phys. Rev. Res.}  \textbf{5}, 033033 (2023).

\bibitem{mumax3} A. Vansteenkiste, J. Leliaert, M. Dvornik, M. Helsen, F. Garcia-Sanchez, and B. Van Waeyenberge, 
\textit{The design and verification
of MuMax3}, AIP Adv. \textbf{4}, 107133 (2014).

\bibitem{metamorphoses} A. O. Leonov, C. Pappas, and I. I. Smalyukh, \textit{Field-driven metamorphoses of isolated skyrmions within the conical
state of cubic helimagnets}, Phys. Rev. B \textbf{104}, 064432 (2021).

\bibitem{Mukai} N. Mukai and A. O. Leonov, \textit{Skyrmion and meron ordering in quasi-two-dimensional chiral magnets}, Phys. Rev. B \textbf{106}, 224428 (2022).

\bibitem{nanomaterials2024} Andrey O. Leonov, \textit{ Reorientation transition between square and hexagonal skyrmion lattices near the saturation into the homogeneous state in quasi-two-dimensional chiral magnets }, Nanomaterials, \textbf{14}, 1970 (2024). 

\bibitem{Butenko2} A. B. Butenko, A. A. Leonov, U. K. Roessler, and A. N. Bogdanov, 
\textit{Stabilization of skyrmion textures by uniaxial distortions in noncentrosymmetric cubic helimagnets}, Phys. Rev. B \textbf{82}, 052403 (2010).


\bibitem{Wieser2017} R. Wieser, R. Shindou, and X. C. Xie, \textit{Manipulation of magnetic skyrmions with a scanning tunneling microscope}, Phys. Rev. B \textbf{95}, 064417 (2017). 

\bibitem{Obadero2020} S. A. Obadero, Y. Yamane, C. A. Akosa, and G. Tatara, \textit{Current-driven nucleation and propagation of antiferromagnetic skyrmionium}, Phys. Rev. B \textbf{102}, 014458  (2017). 

 
\bibitem{Kravchuk2018} V. P. Kravchuk, D. D. Sheka, U. K. Roessler, J. van den Brink, Y. Gaididei, \textit{Spin eigenmodes of magnetic skyrmions and the problem of the effective skyrmion mass}, Phys. Rev. B \textbf{97}, 064403 (2018).


\end{thebibliography}
\end{document}